\documentclass[sigconf]{acmart}
\renewcommand\footnotetextcopyrightpermission[1]{} 
\usepackage{booktabs} 
\usepackage{balance}
\usepackage{filecontents}
\usepackage{xurl}
\usepackage{xcolor}
\usepackage{subfig}
\usepackage{tabularx}
\newcolumntype{Y}{>{\centering\arraybackslash}X}

\usepackage{enumerate}
\usepackage{enumitem}
\usepackage[normalem]{ulem}
\usepackage{float}
\usepackage{multicol}
\usepackage{multirow}
\usepackage{calc}

\usepackage{caption}
\usepackage{stfloats}
\usepackage{makecell}
\usepackage[super]{nth}
\usepackage{cprotect}

\usepackage{tabularx}

\begin{document}
\balance
\title {``Gettr-ing'' Deep Insights from the Social Network Gettr}

\author{Filipo Sharevski}
\affiliation{%
  \institution{DePaul University}
  \streetaddress{243 S Wabash Ave}
  \city{Chicago}
  \state{IL}
  \postcode{60604}
}
\email{fsharevs@cdm.depaul.edu}

\author{Peter Jachim}
\affiliation{%
  \institution{DePaul University}
  \streetaddress{243 S Wabash Ave}
  \city{Chicago}
  \state{IL}
  \postcode{60604}
}
\email{pjachim@depaul.edu}

\author{Emma Pieroni}
\affiliation{%
  \institution{DePaul University}
  \streetaddress{243 S Wabash Ave}
  \city{Chicago}
  \state{IL}
  \postcode{60604}
}
\email{epieroni@depaul.edu}

\author{Amy Devine}
\affiliation{%
  \institution{DePaul University}
  \streetaddress{243 S Wabash Ave}
  \city{Chicago}
  \state{IL}
  \postcode{60604}
}
\email{adevine@depaul.edu}

\renewcommand{\shortauthors}{P. Jachim, F. Sharevski, E. Pieroni, A. Devine}

\begin{abstract}
As yet another alternative social network, Gettr positions itself as the ``marketplace of ideas'' where users should expect the truth to emerge without any administrative censorship. We looked deep inside the platform by analyzing it's structure, a sample of 6.8 million posts, and the responses from a sample of 124 Gettr users we interviewed to see if this actually is the case. Administratively, Gettr makes a deliberate attempt to stifle any external evaluation of the platform as collecting data is marred with unpredictable and abrupt changes in their API. Content-wise, Gettr notably hosts pro-Trump content mixed with conspiracy theories and attacks on the perceived ``left.'' It's social network structure is asymmetric and centered around prominent right-thought leaders, which is characteristic for all alt-platforms. While right-leaning users joined Gettr as a result of a perceived freedom of speech infringement by the mainstream platforms, left-leaning users followed them in numbers as to ``keep up with the misinformation.'' We contextualize these findings by looking into the Gettr's user interface design to provide a comprehensive insight into the incentive structure for joining and competing for the truth on Gettr. 
\end{abstract}



\settopmatter{printacmref=false}
\maketitle
\fancyhead{} 
\pagenumbering{gobble}


\section{Introduction}
Promising a reprieve from banning and content moderation, fringe social networks are offering alternative social media experience to users disenchanted with participation on mainstream places like Twitter or Facebook. One such place is  Gettr (a portmanteau of the words `Get Together'), brandishing an image of an alternative platform founded on ``the principles of free speech, independent thought and rejecting political censorship and `cancel culture''' \cite{Gettr}. This image is not new, as other alternative social networks like Parler offer(ed) users to ``express openly, without fear of being deplatformed for their views'' \cite{Parler}, Gab ``champions free speech'' \cite{Gab}, and 4chan allows ``anyone to post comments and share images'' \cite{4chan}. Therefore, a question arises about what novelty in particular Gettr brings for the fringe communities online.  

Alternative social media platforms attract research curiosity with their lax moderation policies, palpable toxicity, and discourse ridden with overtly polarizing and conspiracy narratives. 4chan, with its notorious, politically incorrect /pol board, receives considerable attention in analyzing trends of trends of self-consciously offensive culture and meme virality on social media \cite{Colley, Mittos, ZannettouC, Hine}. Parler, infamous for providing ``just enough'' networking cohesion for the violent mob attack on the United States Capitol on 6 January 2021 \cite{Munn}, was empirically analyzed to reveal the patterns of amplification of its political pundits and the deliberate user experience design that inhibits a user's ability to search for alternative political narratives \cite{Pieroni, Aliapoulios}. Gab, branded as the ``free speech'' alternative to Twitter, was found to attract alt-right users, conspiracy theorists, and other trolls that disseminate hate speech on the platform  much higher than Twitter, but lower than 4chan's /pol board \cite{ZannettouB, Lima}. And an early look at Gettr \cite{Paudel}, showed yet another outlet for toxicity akin to Gab and 4chan, although yet to achieve the level of engagement and activity characteristic for the online fringe communities. 

These content-focused looks further inspire an important line of inquiry following online extremism \cite{Phadke, Gaudette}, ideological radicalization \cite{Youngblood}, hate speech \cite{Binny, Kennedy}, and false information  \cite{Bleakley}. However, there is a particular absence of two analytical approaches from the the otherwise broader investigation of the `alternative platforms' phenomenon. First, there is virtually no effort to actually gather users' insights from participation on alternative social media. Second, no analysis is concerned about the user experience, and not ideological or expressive, appeal of these platforms. Doing such analyses has nothing to do with legitimizing the platforms' existence or mission, but instead provides a meaningful context to the predominantly content-focused and data-driven investigations so far. The current work, left to conjectures about the future trajectories of the fringe communities online, also falls short of understanding the incentive structure for migrating to, and interacting on, these platforms.

In our study, we weave together a content-driven look at the Gettr, a user study of platform engagement and participation, and a user experience assessment of the Gettr interface in response to this gap. Delving deep into the fabric of the Gettr platform also helped us shed light on the question of the Getter's novelty in the alternative social network space. Contrary to the image of ``independent thoght'' Gettr is largely restrictive of data collection from the platform. Clashing with the the image of ``rejecting the `cancel culture,''' Gettr has \textit{de facto} `canceled' any incentives for anyone but a prominent right-leaning figure to set the tone of the most popular discourses on the platform. The ``free speech'' image is the only one true to the promise, and for that we gathered confirmation form both right-leaning and left-leaning users on the platform, with both group present and engaged in considerable numbers in the daily life on Gettr.

\section{Gettr: Content Analysis}
For our content analysis, we set to collect the data using the GoGettr API, a third party client for scraping data that was created by the Stanford Internet Observatory \cite{GoGettr}. Using the GoGettr API we collected 6.8 million posts, 373,725 users, and 18,274,986 unique follower/followee relationships. The data collection took about two weeks mid-January 2022. To prevent overwhelming Gettr's servers, we added a manual 0.001 second pause after downloading each post, and kept the worker limit well below the threshold referenced in the GoGettr documentation. The dataset includes all posts posted on the platform up to August 10th, 2021, at which point changes to the structure of the database impeded our ability to collect all posts. Rather than switching to using a sample of posts after August 10th, or rather than analyzing the full population as we did before August 10th, we decided to limit the scope of our data collection so that we could analyze a lot of the ``off-topic'' posts that might not be collected using a key word search, or posts by users whose first post was after August 10th. 



\subsection{Initial Data Exploration}

\subsubsection{Post Indexing}
Gettr assigns each post an \verb!_id! using separate base 36 indices beginning with a single letter denoting the type of post being made, e.g. posts start with \verb!p1!, (``Hi \#Gettr''), and comments start with \verb!c1!s. Initially, many of the posts on Gettr uniformly incremented by one, with a couple of missing indices or ``gaps''. Missing indices are to be expected due to deletions, but one gap that we encountered was as high as 200,000. To index the data for our analysis, we started with the very first post we could find, then iterated through the iterator that the GoGettr API provided. Unfortunately, there were frequent errors, and we estimate that in the initial pass, about 60\% of requests resulted in an error. To adjust for this, we created a function that takes the GoGettr API index, makes a prediction of the next most probable one, and queries that post specifically on its own. This filled in about 25\% of the missing posts over the course of multiple passes and we estimate that the final dataset had about 55\% of all posts from that time. 

After approximately August \nth{9}, 2021, the gaps between consecutive posts (posts where the \verb!_id! increased by one) seemed to increase dramatically in both frequency (how often a missing index appeared) and the number of potentially missing indices between the posts (the difference between two \verb!_id!). To confirm that the \verb!_id! no longer increases by one as in prior use, five posts were made in rapid succession from a controlled account.  The post \verb!_id! did not increment by one with each of the five posts but appeared to be randomly generated. Because the \verb!_id! no longer incremented by one, collection using the GoGettr API with our predictive approach was no longer suitable. At this point, we decided to stop collection, analyze the posts that we already had and explore choices for how we might proceed with further data collection.

The way of handling the indices by Gettr most likely does not affect Gettr users, but perhaps changes to the index could be used as a tool to, without causing undue alarm for users, manipulate either the data that people collect or possibly artificially change user metrics. This forced us to perform multiple passes to reliably collect the posts, however, we were wary that the unpredictable nature of the Gettr indexing would still leave out a percentage of the posts on the platform. This makes any data collection methods that depend on the index values impractical. A preliminary conjecture we made at this point was that such an unpredictable behavior could be used to mask content that the Gettr administrators deem off limits for researchers or external data analysis, but do not want to more explicitly ban or hide the content from their users. 

As pointed in \cite{Stanford}, each post/comment object has \textbf{acl} and \textbf{vis} fields which indicate some desire to control/manage the availability of content access. Coupled with the \cite{Stanford} cited administrator control to delete user's posts, a future experiment could be created to explore whether the missing posts prior to August 9, 2022 were from one of these Gettr moderation avenues. Furthermore, inclusion of access control lists and public/private settings does set the platform up for tiers of communication to happen, much like paywalls for online content \cite{Russell}. When Gettr does decide to implement code around the \textbf{acl} and/or \textbf{vis} fields, research could and should be done to evaluate the way communication is monetized on the platform. 

For example, if specific users tend to have unsavory opinions that Gettr administrators do not want to censor, but do not want analyzed by external third parties, incrementing the auto-assigned index a lot before and after they post could make that content less likely to appear in a dataset for possible content analysis using a scraper that relies on the indices.  
Another possible option is that if the index is used to show massive engagement on the platform (which we later discuss in detail in our UX analysis too), then quickly posting and deleting large numbers of posts could be a strategy for artificially increasing the total number of posts being made on Gettr. This, in turn, could skew any metrics or calculations that rely on the total number of posts, e.g. the average number of posts per user. This is not an uncommon behavior on social media as similar attempts has been made in the past to boost the degree to which others perceive an account as influential or powerful \cite{Avram}. 

The unpredictable indexing could also be used as a strategy for curating the data for content analysis. For example, to date, the only paper that analyzes data on Gettr \cite{Paudel} was submitted immediately following the change of from the sequential ids to the random ids (this is obviously a coincidence), but there have not been many more papers written about Gettr since August \nth{9}, 2022 (or not that we are aware of). Additionally, Gettr initially had frictions that made it more difficult for users to post undesirable content like swear words \cite{Stanford}, and the early look of Gettr in \cite{Paudel} noted that as time went on, the unsavory content on the platform was increasing. A pessimistic, but nonetheless plausible, interpretation is that the change to unpredictable indexing might be a deliberate strategy to stifle any systematic review of all content on Gettr from now on. We contacted the GoGettr API team about this abnormal behavior and they acknowledged that they are aware of this sudden change in the indexing and have no explanation or remediation so far.


\subsubsection{Timestamps}
We looked at the \verb!cdate!, the create date associated with a post, for possible frame of reference for content organization. Surprisingly, we uncovered many \verb!cdate! values that preceded the launch date for Gettr -- July \nth{4}, 2021 \cite{mcgraw_team_2021} -- indicating that some posts on Gettr were possibly imported from other platforms, most probably Twitter. The earliest \verb!cdate! we encountered was in 2009 and the \verb!cdate! reached a sharp peak between July \nth{4} 2021 and August \nth{10} 2021, expectedly. Looking at just that peak, we noticed a sharp decrease in the daily number of posts for the first week of July 2021 before the metrics settled into a pattern, and then they seemed to start declining together. 

\begin{figure}[!h]
  \centering
  \includegraphics[width=\linewidth]{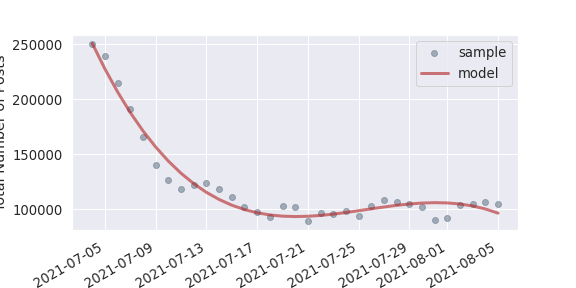}
\cprotect \caption{Regression model showing an apparent decrease in the total number of new unique posts over the first month, using \verb|cdate| as a reference alignment.}
  \label{fig:regression}
\end{figure}

\begin{figure}[!h]
  \centering
  \includegraphics[width=\linewidth]{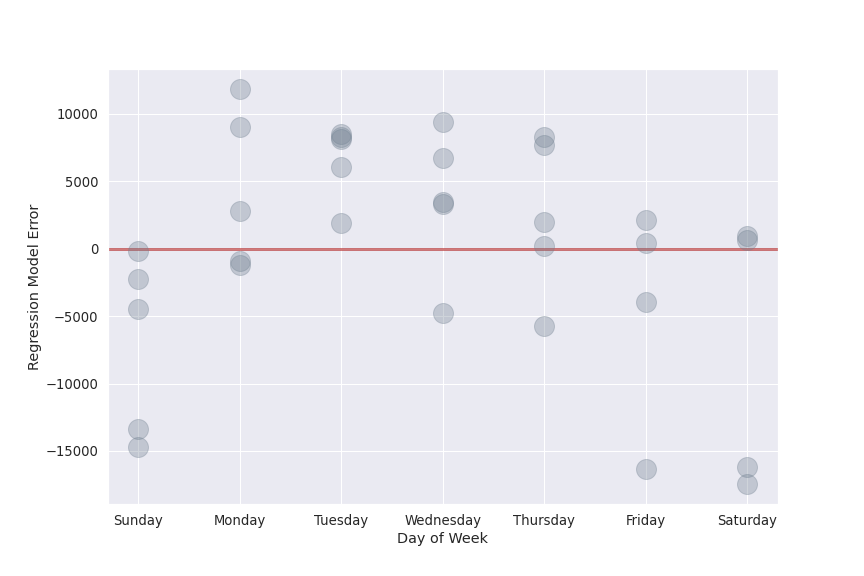}
\cprotect  \caption{Residuals by day of the week.}
  \label{fig:residuals_by_day_of_week}
\end{figure}

We built a third-order univariate regression model using the \texttt{statsmodels} library \cite{statsmodels} that captured 96.3\% of the variability in the total number of posts over that period of time ($r^2$), with an adjusted $r^2$ of 95.9\%. The model is shown in Figure \ref{fig:regression}. To ensure that the success of the model was not due to a high amount of leverage in the last point which could also be an unusually low value, we created several additional univariate third-order regression models dropping the last few points. The adjusted $r^2$ did not fall below 95\%. The residuals show a weekly cyclical pattern, highlighted in Figure \ref{fig:residuals_by_day_of_week}, showing a decrease in platform usage on Fridays, Saturdays, and Sundays.  



Combined with the indices, we proceeded to estimate the actual latest create date or the actual timestamp. For a post at index $i$, we took the max of the estimated date at index $i-1$ and the \verb!cdate! at index $i$ to infer the actual timestamp of the post as shown in Equation \ref{eq:posts}:

\begin{equation}
actual\ timestamp_{i} = \max({date_{i-1}, cdate_i})    
\label{eq:posts}
\end{equation}

\vspace{1em}

Using the actual timestamp, we were able to see if a post's \verb!cdate! was backdated. As an added precaution in case of platforms' idiosyncrasies, like an index being reserved when a draft of a post is created, we looked for posts that appeared to be backdated by at least 7 days. Using this metric, we calculated that 19.68\% or 1,334,891 posts in our sample were backdated at least 7 days. We suspected that many of these backdated posts were not original but imported from platforms similar to the ``public forum'' model of Gettr, such as Twitter, Parler, or even Gab \cite{Hack}. 

\begin{figure}[bhp]
  \centering
  \includegraphics[width=\linewidth]{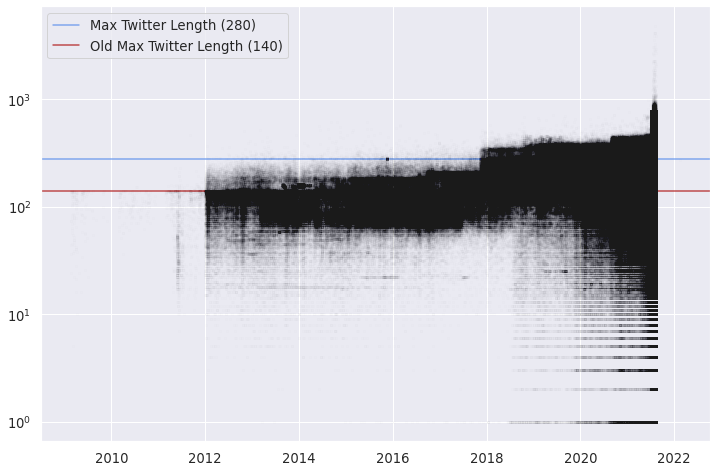}
  \caption{Scatter plot showing the cdate on the x-axis, against the number of characters in the post on a log scale. Note that Gettr was launched on July 4th of 2021.}
  \label{fig:cdate_post_length}
\end{figure}

To test this supposition, we created Figure \ref{fig:cdate_post_length}, in which, we plotted the length of a post (based on the raw count of the characters) on the y-axis against the \verb!cdate! on the x-axis. We found that many of the posts had 140 or fewer characters, until around 2012 when that pattern changed to be about 280 characters. There was also an additional group of posts starting in 2018 with a much higher length in character. These posts could have been imported from Parler, which had a 1,000 character limit, or Gab, with its 3,000 character limit. While the total number of characters in each post did not seem to stop at exactly 1,000 or 3,000, it seems less likely, due to the higher character limit, that users would have to limit their total characters to the same extent that tweets are with their comparatively shorter character limit. 

One possible explanation for the \textit{en masse} backdated import is preservation of content from other platforms that was threatened to be removed or moderated, or, an obscure content that antecedent information for reinforcing existing alternative narratives \cite{Acker}. Twitter and Facebook began to aggressively remove disinformation content after the US 2016 Elections \cite{trollhunterevader} and Parler's entire service was removed in early 2021 \cite{Nicas}, for example. Another possible explanation is to allow the Gettr administrators the opportunity for ``provenance laundering'' of platform information or curating a favorable image to be presented to the Internet public \cite{Xiao}. This is not an unknown phenomenon for social networks, and mainstream platforms' data is never taken to represent human behavior-as-it-is, because of Facebook of Twitter administratively decide to curate data towards maximization of corporate interests \cite{VanDijk}.




\subsection{Linguistic Analysis}
Once we established some basic consistency and reference in the our dataset, we proceeded to do perform a linguistic analysis. Much of the content analysis of the alternative social media sphere so far was focused on meme virality as it comprised the most of the `meme ecosystem' on the Internet \cite{Colley, Mittos, ZannettouC, Hine}. Although this is an equally important line of inquiry, we decided to focus our content analysis towards uncovering linguistic features and use of emojis on Gettr. These two aspects of expression increasingly feature in identity signalling on social media \cite{Roberston}, making a relevant case for being explored on Gettr where the early looks at this platform suggest an increasing level of identity attacks and threats \cite{Paudel}. 

\subsubsection{Matrix Decomposition Analysis}
To look for very similar duplicated posts with similar repeated combinations of keywords, we used a truncated singular value decomposition to analyze the words present in the posts, and used a histogram with the components to look for non-normal distributions, and large numbers of repeated keywords that were frequently used together. This technique could be used to help identify messages that were frequently repeated, or even combinations of terms that were used together with some frequency. To clean the text for analysis, first, we stripped out just the standard latin text from the posts' text, then only kept the terms that appeared in the text at least ten times, keeping text that was longer than one character, and stripping out the stop words using NLTK's default English stop word list (with the addition of the terms ``http,'' ``https,'' and ``:'', all of which indicate a URL) \cite{nltk}. We chose not to lemmatize the terms, because we felt that due to the repetitive nature of some of the vocabulary, and the importance of a lot of the context that's conveyed through the terms' form, that it was most prudent to keep the full words. 


\subsubsection{Cluster Analysis}
Next, we performed a cluster analysis on the posts' text using the same cleaning/preprocessng pipelines that were used for dimensionality reduction. The cluster analysis in this context, uses the frequency of uses of different words with and without each other to find an arbitrary (we landed on 10) groups of posts that share a vocabulary. Using \texttt{KMeans} for clustering, with Term Frequency-Inverse Document Frequency (\texttt{TF-IDF}) for feature vectorization, and single value decomposition to speed up the analysis. The combination of \texttt{TF-IDF} and \texttt{KMeans} is a common pipeline for document clustering, and variations of it have been used in social media contexts \cite{kmeans_social_media_curiskis}. \texttt{TF-IDF} and \texttt{KMeans} has been used to analyze news articles \cite{kmeans_news_shahroz} in conjunction with Single Value Decomposition (although not quite as described in the previous section) \cite{kmeans_news_dimred_kumbhar}, and there are versions of the pipeline that are massively scalable for use on massive distributed systems \cite{kmeans_big_data_yang}. Through a manual review and a trial and error, we decided to use $K = 10$ clusters. The clusters and the number of posts in the training set are given Table \ref{tab:clusters}. We implemented the pipeline using the Scikit-Learn library \cite{sklearn}.

\begin{table*}[!t]
\renewcommand{\arraystretch}{1.0}
\small
\caption{Overview of the clusters created during the cluster analysis showing the algorithmically created clusters of Gettr posts that shared vocabulary, ordered by the number of posts per cluster}
\label{tab:clusters}
\centering
\begin{tabularx}{\linewidth}{|c|l|X|c|}
\Xhline{3\arrayrulewidth}
\textbf{Cluster} & \textbf{Name} & \textbf{Description} & \textbf{Number of Posts}\\\Xhline{3\arrayrulewidth}
1 & Mix Bag Cluster & This cluster contains a mostly random posts. & 2,707,380 \\\hline
2 & Reposts with Description & These are all reposts (akin to retweets in Twitter), but unlike the other retweets in the above cluster, these all include brief descriptions of the post that they are sharing. & 208,410 \\\hline
3 & Pro-Trump & This cluster is dedicated to Donald Trump, showing support in a variety of different ways, including conspiracy theories, concepts surrounding the validity of Biden's victory, references to Q, etc. While many of these posts mention Trump by name, many did not, and were included due to narratives. & 126,310 \\\hline
4 & Reposts & This cluster contains reposts that only have a link, and do not contain any details about the content of the original posts &  11,581 \\\hline
5 & Gettr & These are all posts that mention Gettr in some way. Many of these discuss aspects of the user experience (frustration that more people are not on the platform yet), or reasons for joining the platform (excitement for a real free speech social media experience). & 33,075 \\\hline
6 & Conspiracy & This includes a lot of incendiary stories and spurious connections, much of it seems to be fabricated, as well as connections between different groups (e.g. the LGBT community) and world news. & 29,822 \\\hline
7 & Love & These posts are things that Gettr users ``love'' & 27,007 \\\hline
8 & CCP & Messages that mention CCP, for the Chinese Communist Party. Note, this largely seems to be a stand-in for communist, and may not literally refer to the Chinese Communist Party. & 12,005 \\\hline
9 & Hello &  Posts that say hello to different things. Of these, 19\% have the word ``Gettr,'' others may greet other things like the morning, the world, or ``fellow patriots.'' While this has limited utility from an information warfare perspective, it shows people behaving ``normally'' on the new platform, additionally, when someone says ``hello'' to their ``fellow patriots,'' they are othering people who are not on the platform (or maybe they are referring only to the people on the platform whom they consider to be a ``patriot''). & 9,496 \\\hline
10 & MBC & These posts all involve MBC, short for Middle East Broadcasting Corporation, which is owned by the Saudi government\cite{reuters_mbc}. This shows the presence of government backed news agencies making an effort to have a presence on this platform. & 6,253
\\\Xhline{3\arrayrulewidth}
\end{tabularx}
\end{table*}

Using these clusters, specific types of content on the platform can be identified and isolated. Issues that come up often are LGBT and human rights and the CCP. This means that the CCP could possibly be used as a reason for LGBT rights because of invented stories of how the CCP abuses members of the LGBT community, or inversely, an invented story could say that the CCP protects its LGBT community, therefore, in a free country the LGBT community is oppressed. For example, the total number of posts that reference the Chinese Communist Party could be used as an effective bogeyman for an information campaign or could help the content creator a fast and easy methodology for catching up on the latest conspiracy theories. 


Additionally, the analysis can be re-run with different numbers for the parameter \textit{K} to help label more of the posts that are currently in the ``mix bag cluster,'' and different random starting positions for the starting centroids might yield different clusters that can help show different insights for the content creator. Further, while examining posts within the individual clusters, we identified a few terms and phrases that we may otherwise have missed, that in the context of like posts helped to show continuing trends in how people communicate. A couple of unique terms that we found while using the platform that seemed to have specific meanings included ``TTruth,'' which seems to denote the ``Capital T Truth,'' for a higher level of truth than the truth that they would normally engage with, and ``poop is real,'' which denoted the effects of the ``MSM'' (MainStream Media) in advancing democratic propaganda to blind folks to the TTruth.

\subsection{Emoji Analysis}


Besides the posts themselves, another rich source of text on Gettr are the user account descriptions. One of our early observations is that many user descriptions use lots of emojis. To examine the use of these emojis, we treated the emojis like other text tokens by determining the type of unicode for each emoji we extracted from the scraped Gettr descriptions. We derived 1,406 unique unicode characters. In addition to the emojis, some users chose to write portions of, or even all of their descriptions as special unicode characters. For example, a user might write their description from letters with circles around them. While most people who can read would be able to read those descriptions, the unicode would be stripped from many different text pre-processing pipelines. For example, the TF-IDF default behavior would delete the unicode, leaving only the plain text.  

\begin{figure}[!h]
  \centering
  \includegraphics[width=\linewidth]{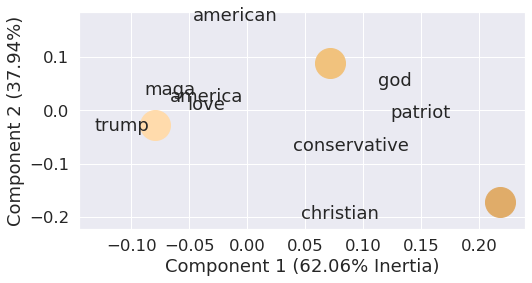}
  \caption{Correspondence analysis with skin tone modifiers and the top 10 words used in descriptions. To interpret the graph, draw an imaginary line from a skin tone to the origin, and perpendicular lines to each of the words, and the order they appear on the first line shows how often they appear together.}
  \label{fig:skin_corr_analysis}
\end{figure}

Another observation that we made is the prevalence of skin tone modifiers in user descriptions. Trying to find ways to target users with different political viewpoints, we leveraged the use of Gettr users' unicode emoji skin tone modifiers to see if the skin tones of the users shows anything about the user's behavior or political alignment. To analyze these effects, we took all descriptions which contained one of the skin tone modifiers, and we plotted a correspondence analysis with the top ten words used in the descriptions, then we ran a Pearson's $\chi^2$ test for the correspondence analysis plotted on Figure \ref{fig:skin_corr_analysis}.

Comparing the presence of the lightest three skin tone modifiers, it looks like the top 10 words that Gettr users use in their descriptions, ``MAGA'' and ``Trump'' are both most closely associated with the palest skin tone. The slightly more tan skin tone is less likely to be associated with Trump, but is more likely to be associated with ``God,'' or ``Patriot.'' The darkest skin tone is more likely to be associated with ``Conservative,'' or ``Christian.'' The $\chi^2$ value for a contingency of these values is 29.26, with 16 degrees of freedom, and a p value of 0.02, indicating that the skin tone does have a relationship with the use of those top terms, with a total of 1,872 samples.These results confirm the evidence of identity signaling \cite{Roberston} by Gettr users through presumably carefully selected words, often describing their political beliefs, and through the skin tone emoji modifier, which is a deliberate choice (as opposed to using the default Simpsonesque yellow).



\subsection{Social Network Analysis}
\begin{figure*}[!t]
\subfloat[This graph shows all friendships on the platform, where users are represented in red dots, and their friendships are represented with black lines between them, or an overlap. The ring of friendships are smaller two or three user friendships subgraphs, and the larger groups in the center are larger subgraphs of friendships.\label{fig:birdseye}]{%
  \includegraphics[width=.45\linewidth]{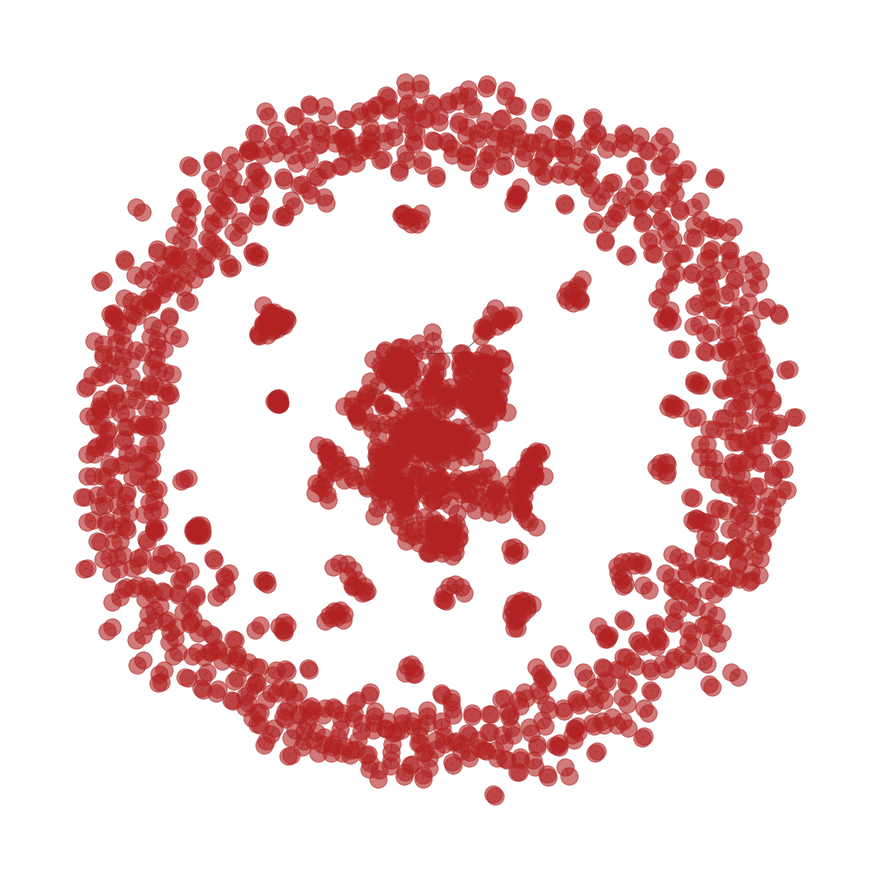}
}\hfill
\subfloat[This is a friendship subgraph of the graph on the left that shows a large number of prominent right-thought leaders, including prominent republicans, republican organizations, and news organizations that tend to favor republicans.\label{fig:right_thinking}]{%
  \includegraphics[width=.45\linewidth]{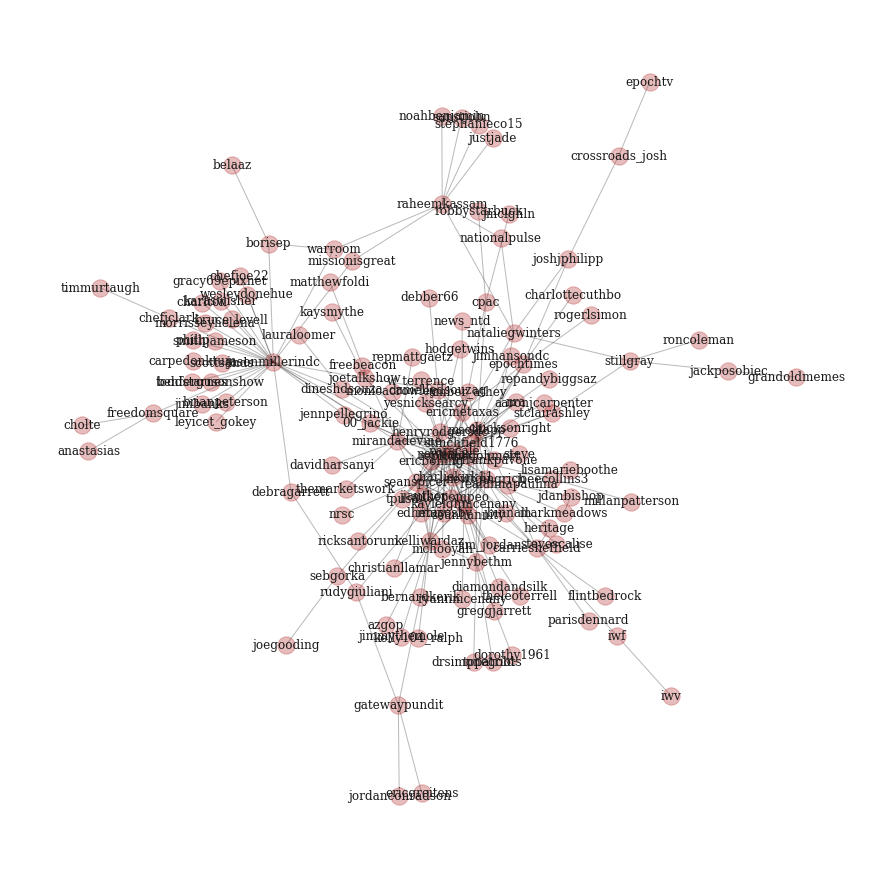}
}
\caption{Friendships on Gettr}
  \label{fig:SPAM}
\end{figure*}

Due to the limitations of the Gettr API to distinguish between different types of follows, we decided to take a look at how often users mention one another, and how frequently two users mention one another. These relationships constitute some form of a  ``friendship,'' as defined in \cite{Huberman_Romero_Wu_2009}, because they show a more deliberate effort between two people to support each other in a manner that is proactive. Previous work on Twitter examining these ``friendships'' has found a stronger positive correlation with a user's engagement on the platform than other relationships with other users like declared follower/followee relationships \cite{Huberman_Romero_Wu_2009}. In our case, due to the relative newness of the platform in the dataset that we created, we limited the number of mentions between two users to be at least one for them to be considered friends. This is simply because users have not had a chance to mention each other too much. We found 1,872 friendship relationships within our dataset, meaning that approximately 0.55\% of users are in a friendship. These friendships naturally formed 592 disconnected subgraphs, of which the majority (500) had just 2 users. A graph of all friendships on the platform is plotted in Figure \ref{fig:birdseye}. 



As we examined the friendship subgraphs, we noted that while most friendship subgraphs only consisted of a couple of users, there were a couple that were substantially larger. The second largest friendship subgraph, depicted in Figure \ref{fig:right_thinking}, consisted of a large number of prominent republicans (Matt Gaetz, Rudy Giuliani, Rick Santorum, Mike Pompeo, Sean Spicer), republican organizations (CPAC, Arizona GOP, the Heritage Foundation), and right-leaning political news organizations (Epoch Times, Steve Bannon's War Room, Newsmax, National Pulse). Together, this shows a public sphere of influence, where people in that subgraph mention one another, and there are clear ideological similarities in the narratives that they present to their audiences.



\section{Gettr: User Insights}
Gathering platform intelligence from Gettr by a content mining and/or user experience analysis is useful in uncovering the macro-level trends that shape the participation on and appeal to this alternative social network. However, a look into Gettr from a user perspective provides us with a context of platform participation on a micro-level, and as such, renders the overall network analysis comprehensive. To our knowledge, this is a first study that brings the individual voices of the Gettr users in fore and analyzes them in the context of the ideological journey through the online fringe communities \cite{Munn}. This perspective provides a macro-level insight into the ``pipeline'' through which users normalize and acclimate to the discourse on the alt-platforms. To this objective we conducted a user study, approved by our Institutional Review Board (IRB) before any research activities began, where Gettr users were invited to expound on: 

\vspace{0.5em}
\begin{enumerate}[label=\Alph*.]
\itemsep 0.5em
    \item What is the value proposition they see in Gettr;
    \item How they participate and express themselves on Gettr;
    \item What kind of information they usually get from Gettr and how they consume it; and
    \item Social media and Gettr experiences. 
\end{enumerate}
\vspace{0.5em}
We set to sample a population that was 18 years or above old, a Gettr account holder, from the United States through the Amazon Mechanical Turk and Prolific. Both reputation and attention checks were included to prevent input from bots and poor responses. The user study asked open-ended questions and it took around 20-30 minutes to complete. Participants were compensated with the standard participation rate of \$27.64/hr. The study was anonymous and allowed users to skip any question they were uncomfortable answering. We also we collected participants' political leanings, race/ethnicity, level of education, gender identity, and age. After the consolidation and consistency checks, a total of 124 participants completed the study. 

The distribution of participants per their self-reported political leanings was: 29  left-leaning (23.38\%), 42 moderate (33.87\%), 40 right-leaning (32.25\%), and 13 apolitical (10.5\%). In respect to race/ethnicity, 91 identified as White (73.38\%), 29 as Black or African American (9.67\%), 13  as Asian (10.48\%), 7 as Latinx (5.64\%), and 1 as Native Hawaiian or Pacific Islander (0.83\%). Education-wise, 22 of the participants had a high-school level (17.74\%) , 84 some college or 2/4-year college (67.74\%), and 19  had a gradate level of education (14.52\%). Gender-wise, 47 of the participants were female (37.9\%), 72 were male (58.06\%), and 6 identified as non-binary (4.04\%). Age-wise, 25  were in the 18 - 24 bracket (20.17\%), 36 in the 25 - 34 (29.03\%), 35  in the 35 - 44 (28.23\%), 16 in the 45 - 54 (12.9\%), and 12 in the 55 - 64 bracket (9.67\%). The distribution of the sample is balanced on the political leanings, gender identity, and age, while skewed towards white and college-level educated participants. 

\subsection{Gettr's Value Proposition}
Our results reveal several reasons why people joined Gettr. The objection to a perceived censorship and freedom of speech infringement by the mainstream platforms, notably Twitter, is prevalent:
\vspace{0.5em}
\begin{description}
\itemsep 0.5em
    \item  [{[P44]}] \textit{Gettr doesn't censor free thought and having been on Twitter for many years, I have seen the decline in the quality of the site. From suspending accounts to slapping false ``fact check'' labels on posts, Twitter has gone from a quality platform to a Gestapo site. Many of my friends have gone over to Gettr and love it, as do I. We aren't free if we can't express thoughts without fear of being banned.} [right-leaning]
    
    \item [{[P38]}] \textit{I don't like the politics of Twitter, and I would love to see an alternative take its place. I am tired of the censorship of conservative voices. I was further annoyed that Parler wasn't allowed to exist (although I am disappointed in Parler's lack of effort to find web hosting.)} [right-leaning]
    
    \item [{[P7]}]\textit{I decided to participate in Gettr because radical feminists are being banned from Twitter left and right just for stating basic facts. I have not been banned but friends of mine have and it's only a matter of time until I am as well.} [left-leaning]
    
    \item [{[P52]}] \textit{The main reason was free speech. A lot of posts on the other social platforms get censored if the discussion gets controversial or it does not fit the narrative of the general populace. It is always important to hear the viewpoints of everyone even if they do not align with yours and hopefully find a common platform where everyone has a voice. } [moderate]
    
    \item [{[P32]}] \textit{I was tired of censorship on other social media platforms. I want to hear different perspectives.} [apolitical]

\end{description}
\vspace{0.5em}
Interestingly, the political leanings across the entire spectrum considerably factored in joining Gettr. The right-leaning and moderate participants were drawn to the freedom of speech value proposition of the platform, while the left-leaning joined Gettr in numbers mostly out of curiosity to see what the `opposition is saying:' 
\vspace{0.5em}
\begin{description}
\itemsep 0.5em
\item [{[P35]}] \textit{Curiosity. I know that it prides itself on not censoring it's subscriber base, so  wanted to see how racist//sexist/transphobic the posts were. I think knowing what the ``other'' side thinks, so as to better inform yourself of the issues and opinions and to know how to combat those ideas.}

\item [{[P20]}] \textit{I wanted to see what the Trump supporters were up to in their new echo chamber.}

\item [{[P48]}] \textit{Although I am quite liberal, I was interested to see what kind of information/misinformation was shared on Gettr. I think there's value in trying to understand the opposing views and trying to learn why they believe the things that they do.}

\item [{[P89]}] \textit{I don't believe or trust anything on that site nor the users and politics that it promotes. I do though want to know what my enemy is up too. My main reason is to observe out of curiosity.}

\end{description}
\vspace{0.5em}

Participants with some high school or high school graduate level were predominantly right-leaning and moderate, indicating that ``\textit{a lot of people [they] follow on twitter have been banned, so [they] joined Gettr to hopefully get uncensored and non biased information again.}'' \textbf{[P71]}. The some college or 2/4-year college graduate participants reflect the aforementioned free-speech/spy-on-conservatives dichotomy as they comprised the largest part of our sample. The post-graduate educated participants, balanced on the political spectrum, provided justifications for what precisely pushed them towards Gettr, for example, ``\textit{Twitter recently they banned Babylon Bee, a satire account for their jokes, which a clear violation of first amendment where's Gettr believes in our constitutional rights.}'' [\textbf{P53}]. Gender-wise, the conformity to the observed dichotomy of the Gettr's appeal is also preserved:

\vspace{0.5em}
\begin{description}
\itemsep 0.5em
\item [{[P75]}] \textit{To try something new for a change to share my ideas about free speech, since like using other platforms such as  Twitter to express said thoughts, was a lot more difficult and hate-arousing.} [male, right-leaning]
\item [{[P12]}] \textit{Curiosity to see what the far-right bubble had to say about things} [male, left-leaning]
\item [{[P37]}] \textit{It seems to be a better source of information than Twitter} [female, right-leaning]
\item [{[P89]}] \textit{I wanted to see what conservatives are posting since i feel like most liberals like myself aren't aware} [female, left-leaning]
\item [{[P101]}] \textit{Some people I follow on twitter created accounts there, so I wanted to see their updates} [non-binary, moderate]
\item [{[P28]}] \textit{I screenshot people saying weird like far-right things and make fun of them with my friends} [non-binary, left-leaning]

\end{description}
\vspace{0.5em}

Age-wise, the participants in the [18 - 24] bracket were mostly joining Gettr for making connections and ``\textit{trying new things, not a fan of the modern social media giants and their censorship}'' [\textbf{P90}]. The [25 - 34] participants added to more context on to the `trying something new' premise, stating that ``\textit{other sites are stale; You deal with the same lack of trust of them and want to get away from all the negativity}'' \textbf{[P108]}. The [35 - 44] participant got even more concrete and stated they joined Gettr ``\textit{so I can talk and converse that have same political beliefs as me and not be ostracized by everyone}'' [\textbf{P59}]. The [45 - 54] participants contextualized this stance by doubling down on the ``\textit{free speech, against the massive push for cancel culture on social media}'' \textbf{[P106]}. The [55 - 64] participants added the differentiation niche with Twitter, as they were ``\textit{was interested to see if indeed Gettr would be better than Twitter}''. \textbf{[P83]}.  

Participants also pointed out that they came to Gettr because ``\textit{some friends of their friends did}]'' [\textbf{P81}] and several noted that ``\textit{many of the conservative talk show personalities that they listen to are on Gettr and they recommended it}'' [\textbf{P122}]. One of the participants found the platform design and features appealing: ``\textit{I enjoy the user interface of Gettr as well as it's more approachable community. As a moderate conservative, I fit in well with almost everyone I have encountered on the platform. Gettr also gives the option to link it to an individual's Twitter.}'' [\textbf{P99}]. The informative value of the content on Gettr was also appealing as \textit{it seems to be a better source of information than Twitter} [\textbf{P37}] where users can ``\textit{stay up to date with news}'' [\textbf{P54}]. 

\subsection{Participation and Expression on Gettr}
Around 27.5\% of all the participants indicated they participate through writing original posts and commenting/liking on other's user posts. Either commenting (29\%) or liking (21.8\%) on other's user posts was how roughly half the participants spent their time on Gettr. The remaining 21.5\% of the participants indicated that they are  ``\textit{mostly just browsing and exploring the platform}'' [\textbf{P29}]. When asked what motivates a user to participate on Gettr, the right-leaning and apolitical participants cited the appeal of a `personally involving discourse' while the moderates and left-leaning cited leaned more towards a `constructive discourse':

\vspace{0.5em}
\begin{description}
\itemsep 0.5em
\item [{[P27]}] \textit{If I have a strong opinion on it or get emotionally worked up. } [right-leaning]

\item [{[P29]}] \textit{I do not fully agree with their conservative agenda, but being on the site and involved in the commentary helps me see the ``other side'' as it was.} [moderate]

\item [{[P86]}] \textit{I am interested in learning more about peoples opinions on many issues, especially political issues that are different than my own.} [left-leaning]

\item [{[P43]}] \textit{If I see interesting topics that resonate with me.} [apolitical]

\end{description}
\vspace{0.5em}

We queried the participants about how Gettr's self-proclaimed `free speech' image facilitates their expression on the platform. The right-leaning participants 
stated that Gettr: ``\textit{doesn't ban people for saying things that big government and the elites might not like}'' [\textbf{P22}], ``\textit{seeks to eliminate the `cancel culture'}'' [\textbf{P33}], ``\textit{Allows free exchange of information that is factual, despite what Twitter says}'' [\textbf{P44}], and ``\textit{allows a person who believes in Donald Trump to be able to express those views and not be censored}'' [\textbf{P59}]. The moderates explicitly highlighted a comparison to Twitter as being ''\textit{strict about removing content that they deem is not correct - even if I don't agree with what someone is saying, I would prefer to see the content instead of having it removed}'' [\textbf{P97}]. The liberal-leaning participants didn't miss to point out that ''\textit{Gettr prides itself on freedom of speech, but obviously it doesn't condone bullying, harassment, threatening behavior, etc; So far concerning `ideas', Gettr is pretty lax and unobtrusive} [\textbf{P35}]. They also pointed out the UX support for freedom of speech, as ''\textit{there is no muting or removing}'' on the platform [\textbf{P36}]. 

In regards to the way of expression, we asked the participants if they use textual content only, emojis, and/or memes (or combination of). The right leaning participants preferred textual expression as not to ``\textit{ `hijack the post' with multimedia} [\textbf{P38}] or to ``\textit{voice different perspectives that the news does not do a good enough job of covering}'' [\textbf{P23}]. The moderates opted for textual expression because ``\textit{it's faster and easier and allows you to be as clear and specific as possible}'' [\textbf{P91}]. Although many liberal-leaning participants come to Gettr to ``\textit{lurk and spy on conservatives}'' [\textbf{P14}], those who actively participate ``\textit{always try to be respectful and still get their point across}'' [\textbf{P34}] as well as use ``\textit{specific wording to balance their views, but not agitate, and shut down discussions} [\textbf{P48}]. 

The right-leaning participants utilized emojis mostly to accentuate a point, e.g. ``\textit{use emojis that correspond with the emotion of my reaction to a post, for example, if I am angry about what is said in a post, I will use the anger emoji} \textbf{[P47]}. The moderates utilized emojis \textit{just to show appreciation for other's posts} \textbf{[P120]}. The left-leaning participants avoided using emojis, arguing that ``\textit{emojis give people a way out, and if I engage, I am generally looking for answers or at least understanding of why certain views are held}'' \textbf{\textbf{[P34]}}. When it comes to memes, the left-leaning and apolitical participants strongly avoided using them, while the moderates used them ``\textit{usually as joke/humor}'' \textbf{[P106]}. Fun was also the most cited reason for using memes by the right-leaning participants, as ``\textit{memes are funny in a political way; I post memes that poke fun at liberals.}'' \textbf{[P122]}. 

\subsection{Information Consumption on Gettr}
Regarding information consumption, we first asked the participants if there is any information they get exclusively on Gettr and nowhere else on social media. Participants singled out: 

\vspace{0.5em}
\begin{description}
\itemsep 0.5em
\item [{[P22]}] \textit{News involving COVID vaccines and statements}

\item [{[P53]}] \textit{Babylon Bee, James O'Keefe, and Project Veritas content and commentary}

\item [{[P59]}] \textit{The MAGA movement information and the ways to de-certify the election}

\item [{[P19]}] \textit{Opinions on minor Republican primary candidates}  
\end{description}
\vspace{0.5em}

We also asked if the participants have compared information between other social networks and Gettr. From the mainstream platforms, expectedly, Twitter was the most sought after place for information comparison across the political spectrum. The left-leaning  and apolitical participants mostly avoided the alternative platforms (less then 8\% in both groups have looked outside of Twitter, Facebook, Instagram or Reddit), while the moderates had a preference for comparison with 4chan. The right-leaning participants were equally interested in comparing Gettr information with Parler, Truth Social, Gab, and 4chan (23.35\%) while maintaining the main focus on the mainstream social networks (76.65\%)


When controlling for gender identity, it is noticeable that male participants have a much stronger preference for comparing information between Gettr on one side, and Reddit, Truth Social, Gab, and 4chan on the other side (35.78\%). The female and non-binary participants mostly turned to the mainstream social networks for information comparison (only 14\% and none, respectively, had looked at any alternative community). The participants with some/high school degree mostly preferred Twitter and Reddit, but also turned to the alternative platforms. The some/college graduates participants had equal preference within the mainstream and alternative platform groups, though much in favor of the Twitter, Facebook, and Reddit (72.64\%). The post-graduate participants mostly sought comparison of information on the mainstream platforms (74.45\%). Age-wise, the trend remains similar to the above, with the less interest for the alternative social networks among the young and more senior participants (only 7\% and 4.5\%, respectively, had looked at any of these communities).

\subsection{Social Media and Gettr Experiences}
We asked the participants if they had had a bad experience on Gettr or another another social network. We did so because our content analysis revealed that 620 unique users mentioned that they were subject of personal harassment in their profile descriptions (a similar trend was observed on Parler where the `banning' was used as a badge of honor \cite{Pieroni}). Citing personal harassment, 10.48\% of the participants reported a bad experience on Twitter, Facebook, Instagram, and Reddit. 

\vspace{0.5em}
\begin{description}
\itemsep 0.5em
\item [{[P123]}] \textit{I've gotten some pretty nasty remarks from people that I don't know on Twitter over content that I did not consider controversial} [right-leaning]

\item [{[P29]}] \textit{I've been shamed as a bigot for doing nothing wrong on Twitter before. I'm not even conservative!} [moderate]

\item [{[P14]}] \textit{Plenty. Discourse with Conservatives on platforms such as Facebook quickly devolves into name-calling and personal attacks} [left-leaning]
\end{description}
\vspace{0.5em}

Our content analysis revealed that 496 unique users mentioned that they got banned from the mainstream social networks into their profile descriptions: 13 of those users explicitly used the term ``Facebook Jail'' badge of honor in their profiles, 32 people self-described themselves as ``shadow-banned'', and \textcolor{red}{...} Citing bans and content moderation, 11.29\% of the participants complained about bad experience on the mainstream platforms:

\vspace{0.5em}
\begin{description}
\itemsep 0.5em
\item [{[P21]}] \textit{I got put in ``Facebook Jail'' a few times for information and thoughts I posted about COVID, which turned me off on using Facebook} [right-leaning] 

\item [{[P30]}] \textit{Yeah, Facebook and Twitter would put stupid warnings on posts about stuff I posted,  but I was actually right} [moderate]

\item [{[P85]}] \textit{I was banned on Facebook for speaking the truth about the \#metoo} [left-leaning]
\end{description}
\vspace{0.5em}

We also asked the participants about what would make them consider leaving Gettr. Interestingly, the dichotomy we observed throughout the analysis so far is somewhat reversed: the left-leaning participants would not leave Gettr for anything while the moderates and right-leaning participants would seek other platforms if censorship/bans crept in on Gettr or users start abandoning it. Drawing on the experience of Parler, some participants expresses worries about Gettr being ``\textit{shut down by Apple or Google} \textbf{[P44]}. An introduction of ``\textit{overwhelmingly biased recommendation algorithm like Twitter}'' \textbf{[P68]}, ``\textit{changes in the UI}'' \textbf{[P79]} and introduction of ``\textit{content moderation}'' \textbf{[P97]} were also mentioned several times besides the main points of free-speech/opposition credo:

\vspace{0.5em}
\begin{description}
\itemsep 0.5em
\item [{[P4]}] \textit{Censorship is the only thing that would cause me to leave. If Gettr ever gets to be like Twitter or Facebook by censoring important stories I will leave to find something better} [right-leaning]

\item [{[P27]}] \textit{If Gettr started banning people and limiting freedom of expression I would probably leave it} [moderate]

\item [{[P14]}] \textit{Nothing, as I only use Gettr for conflict, and to understand the Conservative zeitgeist} [left-leaning]
\end{description}
\vspace{0.5em}

\section{Gettr: User Experience Analysis}

\subsection{Platform Appearance}
In surveying Gettr users on their preferences about the platform, many commented on aspects of the user design/interface that were appealing or impactful based on other platforms, and we thought it fair to highlight some of those here: some users were drawn to Parler based on the user interface itself, calling it ``more approachable'' \textbf{[P99]} than other platforms, as well as ``intuitive'' \textbf{[P41]}, and ``\textit{easily accessible}'' \textbf{[P55]}. The user experience analysis of Parler, the alternative platform preceding Gettr, was characterized by harsh, dark colors and large images which contributed to emotionally charged interaction on the platform \cite{Pieroni}. Parler, in addition, utilized grey and red heavily, with little contrast, that made the platform feel heavy and dreary. Gettr juxtaposes this, with a light, primarily white/light grey color scheme with red accents. The choice of red may be intentional in both cases, which is used to elicit strong emotions, and associated with negative connotations such as threat, danger, and aggression \cite{Elliot}. 

Gettr's logo is characterized by a red torch, often associated with the Statue of Liberty and interpreted as a symbol of freedom within the U.S. in alignment of the launch date of the platform on July 4, 2021, the U.S. Independence Day. On its \textit{About page}, Gettr states its motivations to ``combat censorship and give back freedom of speech to online users'' and overall, these visual cues of freedom are consistent throughout the User Interface (UI). Users seemed to resonate with this message: one user explained that the ``U\textit{I does a much better job of letting its users post whatever they want without the fear of being slapped with a censorship label like on Twitte}r'' \textbf{[P71]}. Another user liked that Gettr doesn't utilize a ``\textit{mute}'' functionality to silence voices on the feed \textbf{[P36]}, and another enjoyed that there is no ``\textit{flag}'' option on posts, to report and potentially remove for misinformation \textbf{[P9]}. These two features work to give users the  \textit{perception} \cite{Stanford}) that ``\textit{there are absolutely no filters when presenting one's viewpoints in the interface}'' \textbf{[P74]}.

\subsection{Profiles}
When navigating to a specific user profile, the layout is familiar to Twitter, with a profile picture and banner, some biographical information, and profile statistics displayed. In reading feedback from Gettr users, popular accounts on platforms like Twitter announced their migration to Gettr and users cited that as a draw to the new platform: ``\textit{Some people I follow on twitter created accounts there, so I wanted to see their updates}'' \textbf{[P10]}. We decided to look at one of the ``celebrity'' profiles on the platform, Sean Hannity (@SeanHannity), a popular night show host to understand the layout of the profile view on Gettr.

\begin{figure}[!h]
  \centering
  \includegraphics[width=0.75\linewidth]{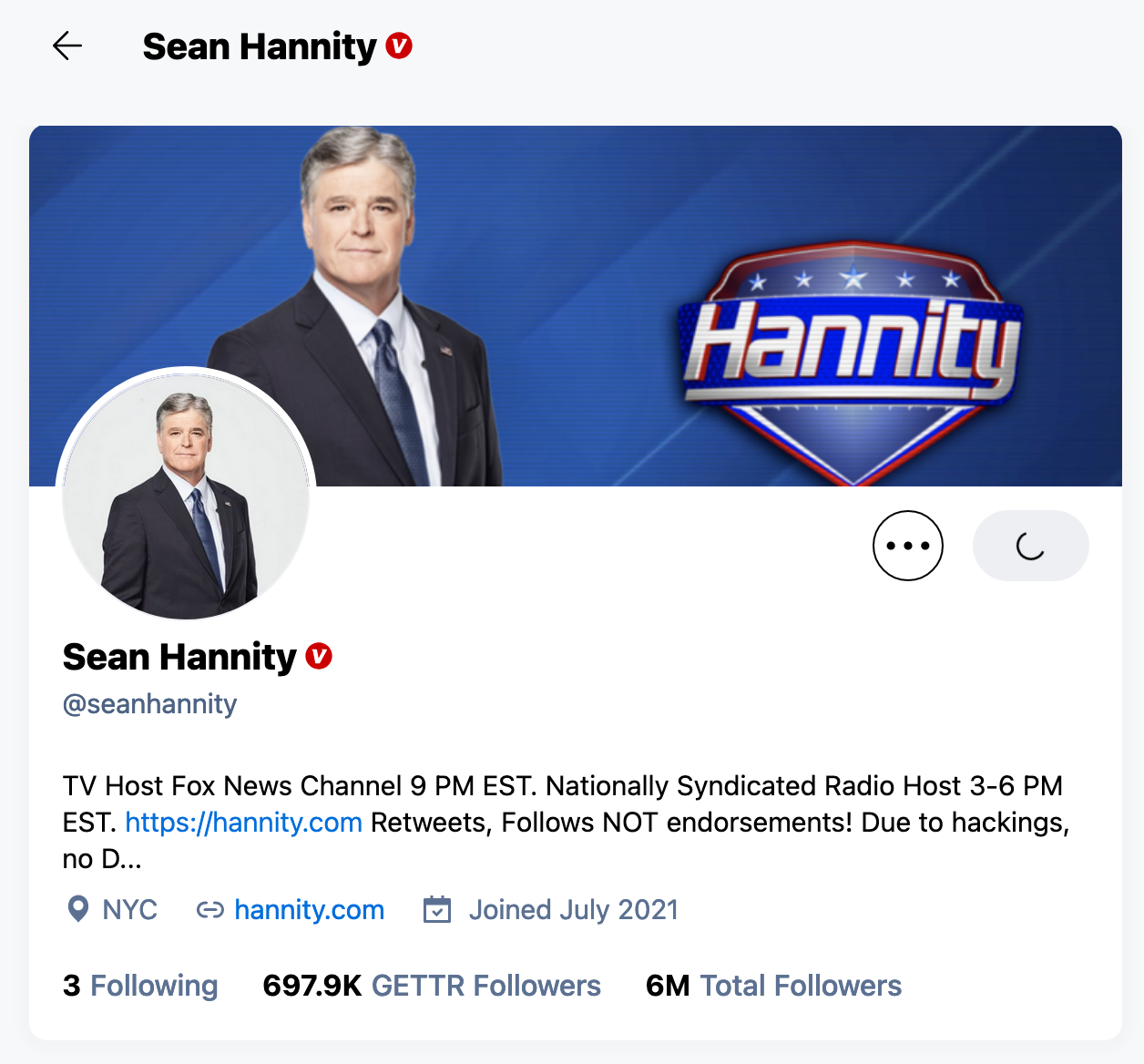}
  \caption{Gettr: Profile Example}
  \label{fig:gettr_profile}
\end{figure}

\begin{figure}[!h]
  \centering
  \includegraphics[width=0.75\linewidth]{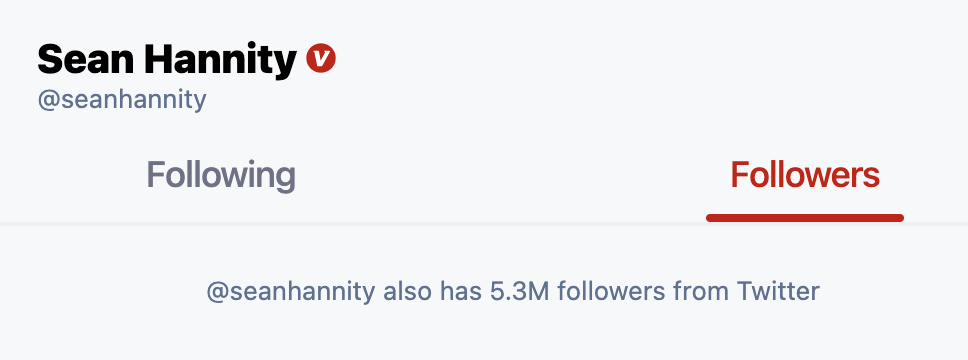}
  \caption{Gettr: Combined Follower Metric with Twitter}
  \label{fig:gettr_followers}
\end{figure}

Curiously, in the case of the profile of Sean Hannity, shown in Figure \ref{fig:gettr_profile}, there are two different follower metrics, ``GETTR Followers'' and ''Total Followers''. The second number  displays a compilation of followers on both Gettr and Twitter. At first, this is unclear, and is only observable after clicking into the ''Total Followers'' view where a disclaimer is observed, shown in Figure \ref{fig:gettr_followers}. On its About page, Gettr claims to offer a reprieve from ``Silicon Valley Mafia’s tyrannical overreach'' yet seems to be allowing its users to profit off of those platforms, by linking their accounts together to amass additional followers and popularity on the alternative platform.

\section{Discussion}
The deep insights from Gettr reveal that users on the platform `get \textit{asymmetrically} together' usually during the work week. Maybe the weekends are left for the other platforms, as virtually everyone we interviewed in the user study confirmed they regularly use other social network, most notably Twitter and Reddit. The asymmetric nature of the discourse becomes a characteristic for the platforms identifying as ``free speech'' guardians as we found a similar centrality towards right-thought ``celebrities'' as is the case on Parler \cite{Pieroni} and Gab \cite{ZannettouB}. Interestingly, we found the discourse of Gettr got a considerable infusion of external posts from other platforms, which could be an indication of either simple duplication so the message of these ``celebrities'' jump-start the platform or the administrators' attempt to create a presentable presentation of the self-appointed ``marketplace of ideas'' branding.

Content-wise, Gettr does not deviate too much from the alt-platform narrative, as it is to be expected with a considerable external post import. Pro-Trump support interspersed with pejorative qualification of left-thought as ``communist'' and linking it with the prominence of the Chinese Communist Party (CCP). The improbable conjectures of political opposition to the conservative agenda were most notably directed towards the LGBT community. This particular alternative narrative is not new and mainstream places attempt to swiftly moderate the misinformation associated with it \cite{Haimson} which turns Gettr in the preferable platform where this alternative narrative shows up again.  

Identity signaling through emoji's is prevalent on Gettr as our results suggest. The Gettr users identifying with MAGA and Trump preferred the palest skin tone, the next skin gradient was mostly preferred by the users identifying through religiousness and patriotism, and the darkest skin gradient was preferred by the users considering themselves conservative and Christian. Perhaps the users on Gettr interpret the default yellow emoji as simply a neutral or unspecified human, none of which is an identity that any of them wants to assume \cite{Roberston}.

We found the evidence for this reasoning in the responses of the Gettr users we sampled to share their experience in participating in the platform. People do bring their identity to the fore on Gettr and virtually it all cases the participation in the discourse encompasses expression of their political attitudes. The ones being moderate or leaning right on the political spectrum cited, in no uncertain terms, their disenchantment of Twitter's banning and moderation as the reason why they joined Gettr. We found more than 10\% of the people in our sample actually being banned or from the mainstream platforms, which is rather considerable number. The ones leaning left, followed in considerable numbers because they saw a value in ``seeing what misinformation the opposition is spreading around.'' 

When it comes to (mis)information, the self-reported topics on Gettr were related to the COVID-19 vaccines and statements, commentary of right-thought pundits and narratives like Project Veritas, MAGA movement information, and republican candidates debates. Interestingly, very few of the participants directly mentioned the topics we identified in the cluster as ones they are interested in or discuss on Gettr. Overall, none of the participants were ready to leave the platform but indicated that it is a possibility if Gettr decides to implement ``censorship''. The moderate and right-leaning participants were also concerned about Gettr being shut down by Apple or Google like Parler was and the left-leaning participants were concerned about where else they can find a similar source of ``conservative zeitgeist.''

From our results it appears that so far users are satisfied with the value proposition of Gettr. We even got far-left participants that found a refuge on Gettr after being banned on Twitter, suggesting that Gettr---at least for now---works to maintain an image they believe is a ``free-speech keeper'' \cite{VoxPop}. Some of the participants expressed that by using this posture Gettr does not condone bullying, harassment, and threatening behaviour, which is an issue that Gettr might face soon or later since the ``marketplace of ideas'' metaphor does  not have static meaning in the broader legal interpretation of the First Amendment online \cite{Schroeder}.



\subsection{Limitations and Future Work} 
Our data collection was limited in the scope of what was provided through existing tools, and the data that was still available on the platform; users deleting posts between when they initially posted and when we scraped them were not included. Additionally, we did not create additional tools for scraping Gettr data that did not rely on the sequentially assigned indexes. As a result, our dataset likely missed a great deal of interesting content, that we would like to write about in the future. There were some reasons that we chose not to use a sampling strategy. Specifically, we considered this to be a unique opportunity to analyze almost all posts created on the platform rather than limiting the analysis to a sample. 

The methods we used were primarily text-driven. This means that messages contained in memes and similar content, which our user study highlighted, is a component in how right-leaning folks communicate, especially with politics and humor. Additionally, some individuals on Gettr use videos in their communications, which we did not review. Many of them also use memes, too, encompassing our future work to focus on multimedia content analysis of the platform as an accompanying part of the textual discourse that captures the emotional aspects of the users \cite{Nissenbaum}. We did however take a preliminary look over the meme and imagery part on Gettr and found out that the most popular meme theme is mockery of the perceived ''left opposition'' while the the most popular motif in the images was the American flag.   

In regards our user study, we were limited in obtaining a larger sample as it was difficult to reach the Gettr user population and many Gettr users declined to participate. Future research, if Gettr is still present in the social network space, could attempt to replicate this study with a larger user population. We also took a very cautious phenomenological approach in surveying the user experience on Gettr. As a next step, we plan to expose Gettr users to particular content or findings about the platform and obtain their opinions as to capture how Gettr maintains (or not) the principles of free speech and independent thought. Another line of inquire following our study is the diffusion of information between Twitter, Reddit, and other alt-platforms that aims to track how the right-thought ``celebrities'' manage to maintain their active presence across the social network sphere as we did not delve deeper to analyze the content of the most popular accounts on Gettr.

\section{Conclusion}
Gettr, at the current stage, appears more as a pathway rather than a fully formed ``pipeline'' through which the disenchantment of the so-called Big-Tech social networks shapes the alternative agenda. With a simple user interface, it does allow acclimation for users perceived as outcasts from the mainstream social media discourse on both sides of the political spectrum and not just on the right. However, we had to uncover this evidence only through user interviews as the Gettr administrators inhibited any effort to objectively determine the larger discourse trends on the platform. We hope our attempt to look deep inside the platform is the first step towards uncovering the `truth' about inner working of Gettr, which, eventually will emerge. Therefore, we call for cooperation rather than \textit{de facto} censorship on the administrative side as we see that of equal benefit for both the Gettr users and evolution of the online social media discourse.

\bibliographystyle{ACM-Reference-Format}
\bibliography{\jobname}

\end{document}